\documentclass{PoS}

\def\mathswitchr#1{\relax\ifmmode{\mathrm{#1}}\else$\mathrm{#1}$\fi}
\def\mathswitch#1{\relax\ifmmode#1\else$#1$\fi}
\newcommand{\PW}{\mathswitchr W}
\newcommand{\PZ}{\mathswitchr Z}
\newcommand{\Pl}{\mathswitch l}
\newcommand{\Pp}{\mathswitchr p}

\newcommand{\X}{X}
\newcommand{\rT}{{\mathrm{T}}}
\newcommand{\MW}{\mathswitch {M_\PW}}
\newcommand{\ri}{{\mathrm{i}}}
\newcommand{\GW}{\mathswitch {\Gamma_\PW}}
\newcommand{\EW}{{\mathrm{EW}}}
\newcommand{\Int}{{\mathrm{IF}}}
\newcommand{\NLO}{{\mathrm{NLO}}}
\newcommand{\veto}{{\mathrm{veto}}}
\def\citere#1{\mbox{Ref.~\cite{#1}}}

\def\de{\delta}
\def\ga{\gamma}

\title{Electroweak corrections to $\PW+\mathrm{jet}$
      hadroproduction including leptonic W-boson decays}

\ShortTitle{EW corrections to $\PW+\mathrm{jet}$ production}

\author{Ansgar Denner\\
        Paul Scherrer Institut, W\"urenlingen und Villigen,\\ 
        Ch-5232 Villigen PSI, Switzerland\\
        E-mail: \email{ansgar.denner@psi.ch}}

\author{Stefan Dittmaier\\
        Albert-Ludwigs-Universit\"at Freiburg, Physikalisches Institut, \\
        D-79104 Freiburg, Germany\\
        E-mail: \email{stefan.dittmaier@physik.uni-freiburg.de}}

\author{Tobias Kasprzik\\
        Max-Planck-Institut f\"ur Physik (Werner-Heisenberg-Institut), \\
        D-80805 M\"unchen, Germany\\
        E-mail: \email{tobiask@mppmu.mpg.de}}

\author{\speaker{Alexander M\"uck}\\
        Paul Scherrer Institut, W\"urenlingen und Villigen,\\ 
        Ch-5232 Villigen PSI, Switzerland\\
        and\\
	RWTH Aachen, Institut f\"ur Theoretische Physik E,\\
	D-52056 Aachen, Germany\\
	E-mail: \email{mueck@physik.rwth-aachen.de}}

\abstract{
This talk summarizes the first calculation of the next-to-leading-order
electroweak corrections to W-boson + jet hadroproduction including
leptonic W-boson decays~\cite{Denner:2009gj}. The W-boson resonance is treated
consistently using the complex-mass scheme, and all off-shell effects
are taken into account. The corresponding next-to-leading-order QCD
corrections have also been recalculated. All the results are
implemented in a flexible Monte Carlo code. Selected numerical results 
for this Standard Model benchmark process are presented for the LHC.
}

\FullConference{The 2009 Europhysics Conference on High Energy Physics,\\
		 July 16 - 22 2009\\
		 Krakow, Poland}

\begin{document}

\section{Introduction}

The production of electroweak (EW) \PW\ and \PZ\ bosons with subsequent
leptonic decays is one of the cleanest and most frequent Standard Model (SM)
processes at the Tevatron and
the LHC. The charged-current Drell--Yan process allows for a precision
measurement of the \PW-boson mass and width, can deliver important 
constraints in the fit of the parton distribution functions, may 
serve as a luminosity monitor at the LHC, and offers the possibility
to search for new charged $\PW{}^\prime$ gauge bosons. For more details
we refer the reader for example to \citere{Gerber:2007xk} and references 
therein, or the numerous talks on physics with EW gauge bosons at 
this conference.

At hadron colliders, the EW gauge bosons are (almost) always
produced together with additional QCD radiation. The production
cross section of \PW\ bosons in association with a hard, visible jet,
\begin{equation}
\Pp\Pp/\Pp\bar\Pp \to \PW  + \mathrm{jet} \to \Pl\nu_\Pl + \mathrm{jet} +\X,
\end{equation}
is still large. The jet recoil can lead to strongly boosted
\PW\ bosons, i.e.\ to events with high $p_{\rT}$ charged leptons and/or 
neutrinos. Hence, $\PW+\mathrm{jet(s)}$ production is not only a 
SM candle process, it is also an important background for a large class 
of new physics searches based on missing transverse momentum. Moreover, 
the process offers the possibility for precision tests concerning jet 
dynamics in QCD.

To match the prospects and importance of this process class, an
excellent theoretical accuracy has already been achieved for the 
prediction of inclusive $\PW$-boson production including NNLO calculations,
resummation, parton-shower matching, NLO EW corrections, and leading 
higher-order corrections. The production of $\PW$ bosons in association 
with jets is now known in NLO QCD up to 3 jets~\cite{Berger:2009ep}. 
An extensive list of references can be found in \citere{Denner:2009gj}.

So far, the EW corrections in the SM have been assessed for 
$\PW+1\,\mathrm{jet}$ production in an on-shell approximation where 
the \PW\ boson is treated as a stable external 
particle~\cite{Kuhn:2007qc}. For \PW\ bosons
at large transverse momentum, i.e.\ at large center-of-mass energy,
this is a good approximation since the EW corrections are dominated by
large universal Sudakov logarithms.

In this work, we present a calculation of the NLO EW
corrections for the physical final state in \PW-boson hadroproduction,
i.e.\ $\Pp\Pp/\Pp\bar\Pp \to \Pl\nu_\Pl + \mathrm{jet} +\X$. 
In contrast to the on-shell approximation, all off-shell
effects due to the finite width of the \PW\ boson are included. Moreover,
we can incorporate the experimental event selection based on the charged-lepton 
momentum and the missing transverse momentum of the neutrino in 
our fully flexible Monte Carlo code
which is able to calculate binned distributions for all physically
relevant $\PW+1\,\mathrm{jet}$ observables. 
Our calculation is completely generic in the sense that it can
predict observables which are dominated by \PW\ bosons close to their
mass shell as well as observables for which the exchanged \PW\ boson
is far off-shell. Moreover, we have recalculated the NLO QCD
corrections at $\mathcal{O}(\alpha^2_\mathrm{s} \alpha^2)$, 
supporting a phase-space dependent choice for the
factorization and renormalization scales.

The calculation of the EW corrections for \PW\ 
production in association with a hard jet is also a step towards a
better understanding of the interplay between QCD and EW corrections
for \PW\ production in general. This understanding---including a full
treatment of off-shell W~bosons---is mandatory to
match the envisaged experimental accuracy for the \PW-mass measurement
at the Tevatron and the LHC.

\section{The Calculation}

In this section we highlight specific aspects of the calculation 
which are particularly important for the presented corrections and
which are not part of the standard framework for NLO corrections. For 
an extensive discussion of the calculational setup we refer the reader 
to \citere{Denner:2009gj}.

The potentially resonant \PW\ bosons require a proper inclusion of the
finite gauge-boson width in the propagators. We use the complex-mass
scheme~\cite{Denner:2005fg}. In this approach the W-boson mass (as well as
the Z-boson mass) is consistently considered as a complex quantity, 
\begin{equation}
\mu_{\PW}^2 = \MW^2 -\ri \MW \GW \, ,
\end{equation}
defined as the location of the propagator pole in the complex plane, where
\MW\ is the conventional real mass and $\GW$ denotes the \PW-boson width.
This leads to complex couplings and, in particular, a complex weak mixing 
angle. The underlying (real) Lagrangian does not change since the 
introduced width is compensated  by adding a corresponding complex 
counterterm. The scheme fully respects all relations that follow from 
gauge invariance. 

The experimental event definition for final-state muons usually selects
so-called ``bare'' muons which are measured without taking into account 
collinear bremsstrahlung photons. Technically, the two collinear particles are
not recombined into a single pseudo-particle and the observable is 
not collinear safe. Therefore, the KLN theorem does not apply and the 
corresponding EW corrections include terms which are enhanced by
logarithms of the (small) muon mass. The enhanced corrections are
phenomenologically relevant and cannot be calculated by the standard subtraction
methods which assume collinear safety. Accordingly, we use an extended
dipole subtraction method~\cite{Dittmaier:2008md} which has been specifically 
designed to deal with non-collinear safe observables. The logarithms are
extracted analytically and we can still work with matrix elements in the
massless muon approximation.

To form collinear-safe quantities, also
photons and QCD partons have to be recombined into a single jet
if they are sufficiently collinear. However, the recombination induces a 
problem if the bremsstrahlung photon and a gluon are accidentally
collinear. In this case, soft gluons can still pass
the jet selection due to the recombination procedure. Hence, a soft-gluon
divergence is induced that would be canceled by the virtual QCD 
corrections to $\PW+\mathrm{photon}$ production. To avoid the singularity,
one has to distinguish $\PW+\mathrm{photon}$ and $\PW+\mathrm{jet}$ 
production by means of a
more precise event definition employing a cut on the maximal energy or
transverse momentum fraction of a photon inside a given jet. However,
this procedure spoils the collinear safety of the event definition in
partonic processes with final-state quarks. Using
again the subtraction formalism~\cite{Dittmaier:2008md} to extract the
problematic collinear terms, the appearance of an unphysical quark-mass 
logarithm in the final result signals the necessity to include
non-perturbative physics to properly describe the emission of a photon
by a quark. The relevant collinear physics can be factorized from the 
underlying hard process and can be cast into a process-independent 
quark-to-photon fragmentation function~\cite{Glover:1993xc}, which has 
been measured at LEP in photon+jet events~\cite{Buskulic:1995au}. We
employ this fragmentation function to achieve both, a realistic event 
selection and a theoretically consistent result.

To reach the accuracy of $\mathcal{O}(\alpha_\mathrm{s} \alpha^3)$
throughout the calculation we have also included the photon-induced
partonic processes and the respective NLO QCD corrections. Also
non-trivial interference terms between EW and QCD diagrams
within the real corrections have been included at this order. However, these
contributions are phenomenologically irrelevant and will not be
discussed in detail in this talk.

\section{Results and Conclusion}
\label{se:results}

We define $\PW+1\,\mathrm{jet}$ events by requiring a jet and a charged lepton 
with transverse momentum $p_{\rT} > 25$~GeV as well as missing transverse
momentum larger than 25~ GeV. The jet and the lepton have to be central with a 
rapidity smaller then 2.5 in absolute value. The details of the event selection
as well as the numerical input values for the calculation can be found in 
\citere{Denner:2009gj}.

\begin{figure}
\includegraphics[width=7.25cm]{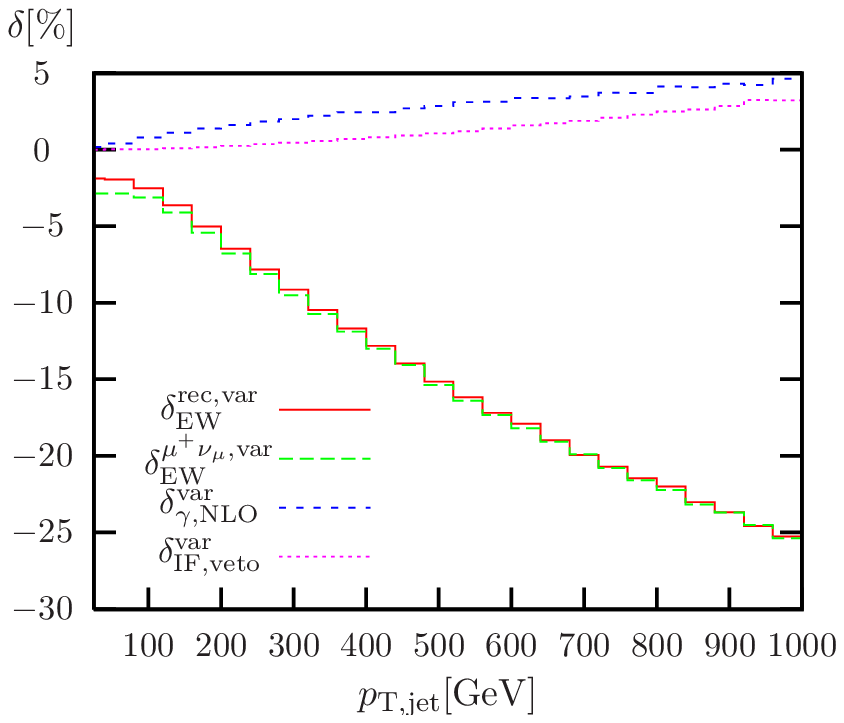}
\hfill
\includegraphics[width=7.25cm]{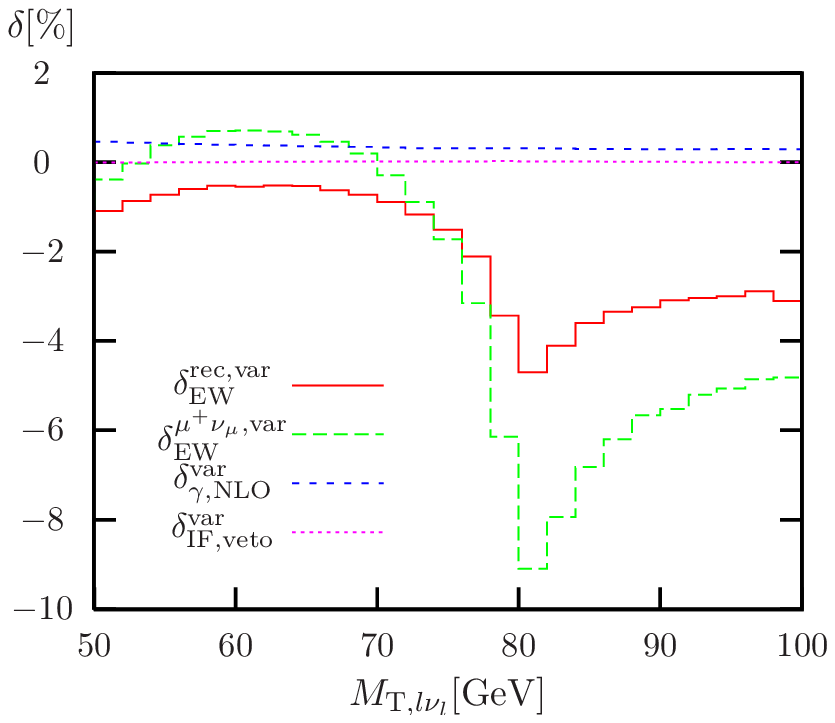}
\caption{\label{fi:results}
Various corrections to the transverse momentum distribution of the 
leading jet (left) and to the transverse-mass distribution of the leptons 
(right). See text for details.}
\end{figure}

For the inclusive cross section, we find negative percent-level EW 
corrections. When we focus on events in the tails of the transverse momentum 
distributions of the charged lepton $p_{\rT,\Pl}$ or the jet 
$p_{\rT,\mathrm{jet}}$ (or the transverse mass distribution of the 
final-state leptons $M_{\rT,\Pl\nu_\Pl}$) we observe the well known 
universal Sudakov enhancement of EW corrections in the high-energy regime.
As shown exemplarily in Figure~\ref{fi:results}, the EW correction rise to 
$-25\%$ at $p_{\rT} = 1$~TeV for the leading jet, both for bare muons 
$\de_{\EW}^{\mu^+\nu_{\mu}\,,\mathrm{var}}$ as well as for lepton--photon 
recombination $\de_{\EW}^{\mathrm{rec}\,,\mathrm{var}}$. 
In the Sudakov regime, where the on-shell result is 
a good approximation, the transverse-momentum distribution for the 
leading jet in Figure~\ref{fi:results} agrees at the percent level with 
the previous on-shell results~\cite{Kuhn:2007qc}. 

For all results 
in this talk we employ a variable scale choice (var) which reflects the 
kinematics of the process and has been chosen to stabilize the QCD 
corrections (see \citere{Denner:2009gj}). Concerning the QCD corrections, 
we only briefly note that a veto against a second hard QCD jet has to be 
used to carefully define the $\PW+1\,\mathrm{jet}$ observable, in particular 
for the $p_{\rT,\mathrm{jet}}$ distribution. Otherwise, the differential 
cross section is completely dominated by QCD dijet production, where a 
quark jet radiates a \PW\ boson, i.e.\ by a completely different process which 
is not related to a generic NLO contribution. 

In Figure~\ref{fi:results}, we also show the small impact of the NLO QCD 
corrected photon induced processes $\de_{\ga,\NLO}^{\mathrm{var}}$ and 
of the interference terms $\de_{\Int,\veto}^{\mathrm{var}}$ for which also 
a sensible jet veto against a second hard jet has been applied.

In contrast to the integrated cross sections, the transverse-mass 
distribution is quite sensitive to the specific
treatment of final-state photons, in particular close to the Jacobian peak of
the distribution at $M_{\rT,\Pl\nu_\Pl}\sim \MW$, where the correction for
bare muons reaches almost $-10\%$  (see 
Figure~\ref{fi:results}). As expected, the corrections for bare muons are 
larger since photons, being radiated collinearly to the charged lepton, 
carry away transverse momentum.

The region around the Jacobian peak is of particular interest for 
the precision determination of the \PW-boson mass. The EW corrections for the
$M_{\rT,\Pl \nu_\Pl}$ distributions resemble the corrections for the
inclusive \PW-boson sample for which no additional jet is required
(see, e.g., Figure~2 in \citere{Brensing:2007qm}). This result is
expected since the transverse mass is rather insensitive to initial-state 
QCD radiation. Our calculation allows to quantitatively compare
distributions for inclusive \PW-boson and $\PW+1\,\mathrm{jet}$ 
production in a future publication. As part of a full NNLO
prediction of the mixed EW and QCD corrections for inclusive  \PW\
production our results can provide a handle for an improved 
understanding of the interplay between EW and QCD corrections in the 
charged-current Drell--Yan process (see \citere{Balossini:2009sa} for a 
recent account of this subject).

To summarize, we have extended the theoretical effort for 
the precise prediction for \PW-boson production at the Tevatron and the
LHC by an important step: We have presented the first calculation of the full 
electroweak (EW) NLO corrections for \PW-boson hadroproduction in association 
with a hard jet where all off-shell effects are taken into account in the 
leptonic \PW-boson decay, i.e.\ we have studied final states with a jet, a
charged lepton, and missing transverse momentum at NLO in the EW coupling
constant within the SM. All results are implemented in a flexible 
Monte Carlo code which can model the experimental event definition at the 
NLO parton level.

\end{document}